\documentstyle[12pt]{article}
\begin{document}
\begin{flushright}
hep-th/0407167\\
SNB/July/2004
\end{flushright}
\vskip 2.5cm
\begin{center}
{\bf \Large { Noncommutativity in the mechanics of \\
a free massless relativistic particle}}

\vskip 2.5cm

{\bf R.P.Malik}
\footnote{ E-mail address: malik@boson.bose.res.in  }\\
{\it S. N. Bose National Centre for Basic Sciences,} \\
{\it Block-JD, Sector-III, Salt Lake, Calcutta-700 098, India} \\

\vskip 1.5cm

\end{center}

\noindent
{\bf Abstract}: 
We show the existence of a noncommutative spacetime structure in the context of
a complete discussion on the underlying spacetime symmetries for the physical
system of a free massless relativistic particle. The above spacetime symmetry
transformations are discussed for the first-order Lagrangian of the system 
where the transformations on the coordinates, velocities and momenta play 
important roles. We discuss the dynamics of this system in 
a systematic manner by exploiting the symplectic structures associated with 
the four dimensional (non-)commutative cotangent 
(i.e. momentum phase) space corresponding to a two dimensional 
(non-)commutative configuration (i.e. target) space. 
A simple connection of the above noncommutativity (NC) is established with the
NC associated with the subject of quantum groups where $SL_{q,q^{-1}} (2)$
transformations play a decisive role.\\

\baselineskip=16pt

\vskip .7cm



\newpage

\noindent
{\bf 1 Introduction}\\

\noindent
The noncommutative geometry and corresponding noncommutative field theories
have generated a great deal of interest during the last few years
due to their clear appearance in the context of string theories and their 
close cousins $D$-branes and $M$-theories (see, e.g., [1-4]). The end points
of the open strings, trapped on the $D$-branes, turn out to be noncommutative
in the presence of the antisymmetric ($B_{\mu\nu} = - B_{\nu\mu}$) potential
$B_{\mu\nu}$ that constitutes the 2-form 
($B = \frac{1}{2!} (dx^\mu \wedge dx^\nu) B_{\mu\nu}$) background 
field $B$. It has also been shown that, in a specific limit, the 
string dynamics can be described as a minimally coupled gauge field theory
defined on a noncommutative space [2].
From a distinctly different perspective, the theoretical 
consideration of the quantum gravity and black hole physics entails upon the 
spacetime to become noncommutative in nature [5,6]. In other words, the NC is
the benchmark of theoretical physics at an energy scale
that is comparable to the Planck energy [6]. Physically, this NC amounts to
the existence of an uncertainty relation between spacetime position operators
which implies that the simultaneous measurement of the spacetime 
positions is not possible to a better accuracy than the Planck length 
(see, e.g., [5-10]).

Even though the NC is connected with the theoretical high energy physics
applicable
at the energy scale of the order of Planck energy, it is expected that the
physical consequences of these NCs can be tested in the low energy 
effective actions for some physically interesting systems. This is why, 
to test the existence of such kind of NC, some  
experimental proposals have been made [11-13] where it has been argued that
only the quantum mechanical effects are good enough to shed some light
on the very existence of the NC in spacetime. There is an alternative
possibility, however. One can construct the low energy theories by exploiting 
the basic ideas behind the NC in spacetime and can propose the physical 
consequences of these NCs on some  experimentally testable 
physical quantities. On the
theoretical side of the latter possibility, mention can be made of the 
noncommutative Chern-Simons theory
[14] and the noncommutative standard model [15] where effects of the NC
have been computed for some interesting physical quantities. However,
experimentally, they are yet to be tested.

It is well-known that the physical insights
and the spacetime symmetries behind the free and/or interacting relativistic
(super) particle are at the heart of the modern developments in the 
understanding of the (super) string theories and (super) gravity
theories. It is also well-understood that a massless scalar relativistic 
particle is endowed with more spacetime symmetries than its massive 
counterpart. In particular, the Lagrangian for the 
free massless relativistic particle respects
the full conformal group of spacetime symmetries which is not 
the case for the free massive relativistic particle
(cf. section 2). In a recent paper
[16], it has been shown that the physical system of a massless 
relativistic particle possesses a new scale type of spacetime symmetry
(in addition to the full conformal group of spacetime symmetries)
which leads
to an extension of the conformal algebra such that the NC in spacetime
emerges naturally (see, e.g., [16] for details). Furthermore, in a recent 
couple of papers [17,18], the reparametrization symmetry and gauge symmetry 
of the {\it massive} relativistic particle have been considered where
the NC in spacetime has been shown to appear
for the specific {\it choices} of the gauge
in the framework of the Dirac bracket formalism. The system of massive
and massless relativistic (super)particles have also been considered
in the framework of quantum groups where the NC has been introduced in
the cotangent (i.e. momentum phase) space of the above physical
system [19-22].

In our present paper, we attempt to study the impact of the NC (in 
the spacetime structure) by laying emphasis on the dynamical
aspects (i.e. equations of motion) as well as the spacetime symmetries
associated with the free massless scalar relativistic particle.
For this purpose, we exploit the symplectic structures in (i) the definition 
of the noncommutative Poisson brackets
(and corresponding commutators), and (ii) the Legendre transformation
to obtain the first-order Lagrangian. In the entire text, we focus on
the symmetry properties of the first-order Lagrangian function together with
the Euler-Lagrange equations of motion that emerge from this function.
The main results of our present endeavour are in three folds. 
First, we attempt to look for the impact of the NC 
on the equations of motion of the free massless relativistic particle. 
Second, we exploit the mathematical sophistication of symplectic structures
to study the details of the commutative and NC dynamics. Third, we establish
a connection between (i) the NC of spacetime that emerges from the scale type
of spacetime symmetry, and (ii) the NC of spacetime that originates from the
quantum group $GL_{q,q^{-1}}(2)$ (and its special case
$SL_{q,q^{-1}}(2)$) symmetry. It should be noted that all these results
are {\it valid} only up to the lowest order (i.e. $\sim \beta = e^2 p^2$) 
in the parameters of the scale symmetry transformations that lead to the
existence of NC in the theory.

In the context of the above results, it is pertinent to
point out that, by thorough discussions on the key mathematical aspects of
the dynamics, we find that the equations of motion remain unaffected
due to the presence of the NC generated by
the new scale type of spacetime symmetry in the theory. 
This observation is true to the lowest  order in the parameters 
($\sim \beta = e^2 p^2$) of the above scale type of spacetime 
symmetry transformation.
It should be noted that, even though the NC exists in the commutators
to the lowest order ($\sim \beta = e^2 p^2$, cf. (3.6)), the equations of
motion (cf. (4.10), (4.32)) and the Hamiltonian (cf. footnote just after (4.32))
remain unaffected due to this NC because of the fact that all the terms, linear
in the parameters ($\sim \beta = e^2 p^2$), cancel out. 
Furthermore, we have devoted a great deal of
discussions on the derivation of the components of the 
covariant symplectic metric so that the equations of motion for the canonical
{\it commutative} case as well as the nontrivial 
{\it noncommutative}  case could be contrasted against each-other
with utmost accuracy 
(cf. section 4). However, as it turns out, the equations of motion remain
unchanged even though the parameters of NC (linear in $\sim \beta = e^2 p^2$)
are present
in the covariant symplectic structure
(cf. (4.29)) on the noncommutative cotangent manifold.

We have been able to demonstrate a  connection of the NC associated with
the new scale type spacetime transformations to the NC associated with the
quantum group $GL_{q, q^{-1}} (2)$ transformations on the phase variables
where the elements of this group have been chosen in such a way
that it automatically
becomes $SL_{q,q^{-1}} (2)$. It turns out that the noncommutative Poisson 
brackets between
phase variables due to the $SL_{q, q^{-1}} (2)$ symmetry
transformations reduce to the noncommutative  Poisson brackets due to 
new scale type spacetime
symmetry transformations for the deformation parameter satisfying $q^2 = 1$.
Furthermore, the choice of the elements of $GL_{q, q^{-1}} (2)$ are such
that the key $q$-algebraic relations among the phase variables [22] of
the free massless relativistic particle
remain intact on the cotangent manifold. To be more precise, 
for $q^2 = 1$, the relationships 
among the phase variables (that respect Lorentz invariance
and $SL_{q,q^{-1}} (2)$ invariance) remain unchanged.

The contents of our present paper are organized as follows. In section 2,
we set up the notations and conventions by discussing the bare essentials
of the conformal, gauge- and reparametrization symmetry transformations
for the Lagrangian of the free massless relativistic particle. For
the paper to be self-contained, section 3 deals, in a somewhat different 
manner, the NC in the spacetime structure that owes its origin to 
an additional scale symmetry transformation [16].
Section 4 is devoted to the discussion of dynamics
for the free massless relativistic particle in the four-dimensional 
(non-)commutative cotangent manifold. In section 5, we demonstrate a 
simple connection between the NC of spacetime due to 
an additional scale symmetry 
and the NC of spacetime due to quantum group 
$SL_{q,q^{-1}}(2)$ symmetry. Finally, we make
some concluding remarks in section 6 and point out a few future directions
for further investigations in the connected areas of research.\\

\noindent
{\bf 2 Preliminary: symmetries in Lagrangian formalism}\\

\noindent
We begin with the different looking but equivalent forms of the 
gauge- and reparametrization
invariant Lagrangians for a free massive relativistic particle. The particle
moves on a world-line (i.e. trajectory) that is embedded in a $N$-dimensional
flat Euclidean target manifold
\footnote{For the Euclidean target manifold,
we choose the metric $\delta_{\mu\nu} = $ diag $(+ 1, + 1, ......+1)$
and the scalar product between two vectors $A_\mu$ and $B_\mu$
is given by $(A \cdot B) = \delta_{\mu\nu} A_\mu B_\nu$. Thus,
the contravariant vectors are same as the covariant vectors. In other
words $A_\mu B^\mu = A_\mu B_\mu  
= A_1 B_1 + A_2 B_2 +....+ A_N B_N $. For the sake of
convenience, however, we shall be using the upper and lower indices
in the whole body of our text.}.  
The specific Lagrangians, describing the above massive particle, are [23,24]
$$
\begin{array}{lcl}
L^{(m)}_{0} = m \; (\dot x^2)^{1/2} \;\quad
L^{(m)}_{f} = p_\mu \dot x^\mu - \frac{1}{2}\;e\; (p^2 - m^2) \;\quad
L^{(m)}_{s} = \frac{1}{2}\; e^{-1}\; (\dot x)^2 + \frac{1}{2}\; e \; m^2.
\end{array} \eqno(2.1)
$$
In the above, the mass-shell condition ($p^2 - m^2 = 0$) and the force free
(i.e. $\dot p_\mu = 0$) motion of the {\it free} massive relativistic particle
are a couple  of  common 
features for (i) the Lagrangian with the square root $L^{(m)}_{0}$, (ii) the
first-order Lagrangian $L^{(m)}_f$, and (iii)
the second-order Lagrangian $L^{(m)}_s$. Except for the  mass 
(i.e. the analogue of 
the cosmological constant) parameter $m$, the target space canonically 
conjugate coordinates $x^\mu (\tau)$
(with $\mu = 1, 2......N$) as well as the momenta $p_\mu (\tau)$ and
the einbein field $e (\tau)$ are the functions of the monotonically increasing
parameter $\tau$ that characterizes the trajectory (i.e. the world-line)
of the free massive scalar 
relativistic particle. Here $\dot x^\mu = (d x^\mu/d \tau)$ are the
generalized versions of the
``velocity'' of the particle. The first- and the second-order Lagrangians 
are  endowed with the first-class constraints $\Pi_e \approx 0$
and $p^2 - m^2 \approx 0$ in the language of the Dirac's classification 
scheme where $\Pi_e$ is the canonical conjugate
momentum corresponding to the einbein field $e(\tau)$. The existence of
the first-class constraints on this physical system
establishes the fact that this reparametrization
invariant theory of the free massive 
relativistic particle is a {\it gauge} theory
\footnote{The Lagrangian density $L^{(m)}_f$ transforms to
$\delta_r L^{(m)}_f = (d/d \tau) [( \epsilon L^{(m)}_f)]$ under the 
transformations
$\delta_r x_\mu = \epsilon \dot x_\mu, 
\delta_r p_\mu = \epsilon \dot p_\mu, 
\delta_r e = (d/d\tau) [(\epsilon e)]$ generated by the 
basic reparametrization $\tau \to
\tau - \epsilon (\tau)$ where $\epsilon (\tau)$ is an infinitesimal parameter. 
Similarly, under the gauge
transformations $\delta_g x_\mu = \xi p_\mu, \delta_g p_\mu = 0, \delta_g e
= \dot \xi $, the Lagrangian density $L^{(m)}_f$ transforms to a 
total derivative.
Both these transformations are equivalent (with the identification
$\xi = e \epsilon$) for the free (i.e. $\dot p_\mu = 0$) 
relativistic particle because
both the above transformations owe their origin to the mass-shall condition
$p^2 - m^2 = 0$. Thus, conditions $\dot p_\mu =0$ and $p^2 -m^2 = 0$ are
a couple of salient features for the dynamical description
of the relativistic particle.}. It is clear that the massless limit
(i.e. $m \to 0$) is not consistently defined
 for the Lagrangian $L^{(m)}_0$ but the first-
and the second-order Lagrangians do permit such a limit. The Lagrangians
for the massless free relativistic particle, derived in such a limit
(i.e. $m \to 0$), from $L_f^{(m)}$ and $L^{(m)}_s$:
$$
\begin{array}{lcl}
L_{f} = p_\mu \dot x^\mu - \frac{1}{2}\;e\; p^2 \;\;\;\;\;\qquad\;\;\;\;
L_{s} = \frac{1}{2}\; e^{-1}\; (\dot x)^2
\end{array} \eqno(2.2)
$$
are {\it not only} endowed with the following Poincar{\'e} ($\delta_p$),
reparametrization ($\delta_r$) and gauge  ($\delta_g$) symmetry transformations
which are also present for their massive counterparts (2.1)
but they also respect scale ($\delta_s$) and conformal ($\delta_c$)
symmetry transformations that are {\it not}
present for the Lagrangians (2.1) for the
massive free relativistic particle. In more sophisticated language, the 
breaking of the latter symmetries is said to generate the mass of the
free particle. It is interesting to check that
the second order Lagrangian $L_s$ of (2.2)
respects the following symmetry transformations 
$$
\begin{array}{lcl} 
&&\delta_g x_\mu = \xi \bigl ( {\displaystyle \frac{\dot x_\mu}{e}} \bigr )\;
\qquad \;\;\;\delta_g e = \dot \xi\; \qquad \;\;\;\;\delta_g L_s =
{\displaystyle \frac{d} {d\tau} \Bigl [ \frac{\xi}{2}\; \frac{\dot x^2}{e^2}
\Bigr ]} \nonumber\\
&& \delta_s x_\mu = \alpha x_\mu\; \;\;\;\qquad \;\;\;
\delta_s e = 2 \alpha e\; \;\;\;\qquad \;\;\;\delta_s L_s = 0 \nonumber\\
&& \delta_c x_\mu = 2 x_\mu (x \cdot b) - b_\mu x^2\; \qquad
\delta_c e = 4 e (x \cdot b)\; \qquad \delta_c L_s = 0 \nonumber\\
&& \delta_p x_\mu = \omega_{\mu}^{\nu} x_\nu + a_\mu\; \;\;\;\qquad\;\;\;
\delta_p e = 0\;\;\;\; \qquad \;\;\;\delta_p L_s = 0 \nonumber\\
&&\delta_r x_\mu = \epsilon \dot x_\mu\; \;\;\qquad\;
\delta_r e = 
{\displaystyle \frac{d}{d \tau}} \bigl [ \epsilon e \bigr ]\; 
\;\;\qquad \;\delta_r L_s = 
{\displaystyle \frac{d}{d \tau}} \bigl [ \epsilon L_s \bigr ] 
\end{array} \eqno(2.3)
$$
where $\xi (\tau), \epsilon (\tau)$ are the {\it local}
 infinitesimal parameters
corresponding to the gauge and reparametrization symmetry transformations,
respectively,
and $\omega^{\mu\nu}$ (with $\omega^{\mu\nu} = - \omega^{\nu\mu}$),
$a^\mu$, $\alpha$ and $b^\mu$ are the {\it global} infinitesimal parameters
corresponding to the Poincar{\'e} transformations ($\omega, a)$, 
scale transformation ($\alpha$) and conformal transformations ($b^\mu$), 
respectively. The same symmetry transformations,
for the first-order Lagrangian $L_f$, are
$$
\begin{array}{lcl} 
&&\tilde \delta_g x_\mu = \xi p_\mu\; \qquad \tilde \delta_g p_\mu = 0\; \qquad
\qquad \tilde \delta_g e = \dot \xi\; \qquad \tilde \delta_g L_f =
{\displaystyle \frac{d} {d\tau} \Bigl [ \frac{\xi}{2}\; p^2 \Bigr ]}
\nonumber\\ 
&&\tilde \delta_s x_\mu = \alpha x_\mu\; \qquad 
\tilde \delta_s p_\mu = - \alpha p_\mu\; \qquad
\tilde \delta_s e = 2 \alpha e\; \qquad \tilde \delta_s L_f = 0 \nonumber\\
&& \tilde \delta_c x_\mu = 2 x_\mu (x \cdot b) - b_\mu x^2\; \;\;\;\qquad\;\;
 \tilde \delta_c e\; =\; 4 e (x \cdot b) \nonumber\\
&& \tilde \delta_c p_\mu =  2 x_\mu (p \cdot b) - 2 b_\mu (x \cdot p)
- 2 p_\mu (x \cdot b)\; \;\;\;\;\;\qquad \;\;\;\;\;
\tilde \delta_c L_f = 0 \nonumber\\
&& \tilde \delta_p x_\mu = \omega_{\mu}^{\nu} x_\nu + a_\mu\; \qquad
\tilde \delta_p p_\mu = \omega_\mu^\nu p_\nu\; \qquad
\tilde \delta_p e = 0\; \qquad \tilde \delta_p L_f = 0 \nonumber\\
&& \tilde \delta_r x_\mu = \epsilon \dot x_\mu\; \qquad
\tilde \delta_r p_\mu = \epsilon \dot p_\mu\; \qquad
\tilde \delta_r e =
{\displaystyle \frac{d}{d \tau}} \bigl [ \epsilon e \bigr ] 
\qquad \tilde \delta_r L_f = 
{\displaystyle \frac{d}{d \tau}} \bigl [ \epsilon L_f \bigr ]. 
\end{array} \eqno(2.4)
$$
It will be noted that all the above symmetry transformations for the
momentum variable $p_\mu$, present in the first-order
Lagrangian $L_f$, are such that there is a mutual consistency among
the transformations on $x_\mu, e$ and $p_\mu$ so that the
sanctity of the relation
$p_\mu = e^{-1}\; \dot x_\mu$, derived from $L_s$ or $L_f$, 
could be maintained. As explained earlier, the generator of the 
reparametrization transformation $\delta_r$ and the gauge transformation
$\delta_g$ is the mass-shell condition $p^2 - m^2 = 0$ and
$p^2 = 0$ for the massive- and massless cases, respectively. The operator
form of the generators of all the above ``conformal'' transformations are 
$$
\begin{array}{lcl} 
&&\hat P_\mu = \partial_\mu\; \qquad 
\hat K_\mu = (2 x_\mu x^\nu - x^2 \delta_\mu^\nu)\;
\partial_\nu \nonumber\\
&& \hat D = x^\mu \partial_\mu\; \qquad \hat M_{\mu\nu} = x_\mu \partial_\nu
- x_\nu \partial_\mu
\end{array} \eqno(2.5)
$$
where the angular momentum operator
$\hat M_{\mu\nu}$ and momentum operator $\hat P_\mu$ are the generators of the
rotation and translation  that constitute the full
Poincar{\'e} transformations. The scale transformation $\delta_s$
is generated by the dilation operator $\hat D$ and the conformal
transformation $\delta_c$ is generated by the conformal boost
operator $\hat K_\mu$ (see, e.g., [16] for more details). \\

\noindent
{\bf 3 Additional symmetry and noncommutativity}\\

\noindent
It is clear from the starting
Lagrangian $L^{(m)}_0$ in (2.1) that the Euler-Lagrange equations
of motion are: $\ddot x_\mu  (\dot x)^2 
- \dot x_\mu (\dot x \cdot \ddot x) = 0$.
For the free motion $\ddot x^\mu = 0$ of the particle, we can
choose a gauge such that $(\dot x \cdot \ddot x) = 0$ 
and $(\dot x)^2 \neq 0$ (see, e.g., [23]).
It is straightforward to notice that the mass $m$ of
the particle does not play any role at all in the dynamics of the particle.
Thus, the first-order- and the second-order Lagrangians of (2.1) and
(2.2) produce exactly the same type of equations of motion. These
are encompassed in the free motion ($\dot p_\mu = 0$) of the particle
and the definition of the canonical momentum 
(i.e. $ p_\mu = e^{-1} \;\dot x_\mu$). In other words, the combination
of these two relationships yields the following Euler-Lagrange 
equation of motion from the
first- and the second-order Lagrangians
$$
\begin{array}{lcl} 
\ddot x_\mu e - \dot x_\mu \dot e = 0 \;\;\;\Rightarrow\;\;\;
(\dot x \cdot \ddot x) e - (\dot x)^2 \dot e = 0.
\end{array} \eqno(3.1)
$$
For the free motion $\ddot x_\mu = 0$ of the particle, we have to choose
the following gauges: $(\dot x \cdot \ddot x) = 0$ and $\dot e = 0$
(but $ e \neq 0$). There
are a few compelling reasons for the above choices. First, in the
limit ($ e \to 0$), we should recover the gauge choice imposed on the
equations of motion derived from $L^{(m)}_0$ to obtain the free motion
(i.e. $\ddot x_\mu = 0$). Second, the einbein field $e (\tau)$ is like
the gauge field $A_\mu$ of the Abelian 1-form gauge theory where the
Lorentz gauge $(\partial \cdot A) = 0$ is just the analogue of $\dot e = 0$.
In fact, this gauge choice is exploited in the Becchi-Rouet-Stora-Tyutin
(BRST) quantization (see, e.g., [21,24])
of the free massive as well as massless relativistic particle
\footnote{In fact, the (anti-)BRST invariant Lagrangian 
$L_b = p_\mu \dot x^\mu - \frac{1}{2} e (p^2 - m^2) + b \dot e 
+ \frac{1}{2} b^2 - i \dot {\bar c} \dot c$ does exploit the gauge-fixing term
$\dot e$ through the Nakanishi-Lautrup auxiliary field $b$. In terms of the
(anti-)ghost fields $(\bar c)c$, the nilpotent ($s_{(a)b}^2 = 0$)
(anti-)BRST symmetry transformations $s_{(a)b}$ are:
$s_b x_\mu = c p_\mu, s_b p_\mu = 0, s_b c = 0, 
s_b e = \dot c, s_b \bar c = i b, s_b b = 0$
and $s_{ab} x_\mu = \bar c p_\mu, s_{ab} p_\mu = 0, s_{ab} \bar c = 0,
s_{ab} e = \dot {\bar c},
s_{ab} c = - i b, s_{ab} b = 0$ [21,24].}.
Third, it is evident that the constraint $ p^2 \approx 0$ is the generator
of the gauge- and reparametrization symmetries (cf. (2.4)) for the massless
relativistic particle. Exploiting $  p_\mu = e^{-1} \dot x_\mu$, it can
be seen that the time invariance of the above constraint
$(d /d \tau) [\; p^2 \;] = 0$ (which is equivalent to
$( p \cdot \dot p) = 0$) implies that the gauge choices $(\dot x \cdot
\ddot x) = 0$ and $\dot e = 0$ are consistent with the free motion
$\ddot x_\mu = 0$ of the massless relativistic particle. Utilizing
the free motion $\dot p_\mu = 0$ and the above gauge choices
(i.e. $\dot e = 0$ and $(\dot x \cdot \ddot x) = 0)$, it is
straightforward to check that the following scale transformations
on the spacetime coordinate $x_\mu$, momentum $p_\mu$
and einbein field $e (\tau)$
$$
\begin{array}{lcl}
x_\mu (\tau) \rightarrow X_\mu (\tau) &=& 
{\displaystyle e^{\beta (\dot x^2)}}\; x_\mu (\tau)
\equiv 
{\displaystyle e^{\beta (e^2 p^2)}}\; x_\mu (\tau) \nonumber\\ 
p_\mu (\tau) \rightarrow P_\mu (\tau) &=& 
{\displaystyle e^{- \beta (\dot x^2)}}\; p_\mu (\tau)
\equiv e^{- \beta (e^2 p^2)}\; p_\mu (\tau) \nonumber\\ 
 e (\tau) \rightarrow E (\tau)  &=&  e^{2 \beta (\dot x^2)}\; e (\tau)
\equiv e^{2 \beta (e^2 p^2)}\; e(\tau) 
\end{array} \eqno(3.2)
$$
leave the  first-order Lagrangian $L_f$ invariant primarily because
of the fact that the above transformations imply: $\dot x^\mu \to \dot X^\mu 
= e^{\beta(e^2 p^2)} \dot x^\mu$. This is
an extension, albeit in a restricted sense, of the conformal symmetries
listed in (2.3) or (2.4). As a consequence, the dilatation operator
$\hat D = x^\mu \partial_\mu$ is extended to $\hat D^* = (1 + \beta) \hat D$.
Furthermore, the above new symmetry allows
an extension of the conformal algebra which has been discussed 
thoroughly in [16]. The infinitesimal version 
(i.e. $ \beta^n \approx  0, n \geq 2$) of (3.2),
which are of physical importance in spacetime symmetries, are 
$$
\begin{array}{lcl}
x_\mu (\tau) \rightarrow X_\mu (\tau) &=& x_\mu (\tau) +
\beta (e^2 p^2)\; x_\mu (\tau) \nonumber\\ 
p_\mu (\tau) \rightarrow P_\mu (\tau) &=& p_\mu (\tau) 
- \beta (e^2 p^2)\; p_\mu (\tau) \nonumber\\ 
e (\tau) \rightarrow E (\tau)  &=& e (\tau)
+ 2 \;\beta (e^2 p^2)\; e(\tau). 
\end{array} \eqno(3.3)
$$
It is now clear that the following canonical brackets
$$
\begin{array}{lcl}
[x_\mu, x_\nu] = 0\; \quad 
[p_\mu, p_\nu] = 0\; \quad [x_\mu, p_\nu] = i \delta_{\mu\nu}\;
\quad  [x_\mu, e] = 0\;\quad [p_\mu, e] = 0
\end{array}\eqno (3.4)
$$
emerging from the first- and the second-order Lagrangians 
of the massless relativistic particle, are now
changed to their noncommutative counterparts as 
$$
\begin{array}{lcl} 
&& \bigl [ X_\mu (\tau), X_\nu (\tau) \bigr ]\; = \;
(1 + \beta)\; \Bigl \{ [x_\mu (\tau), \beta]\; x_\nu (\tau)
+ [ \beta, x_\nu (\tau) ]\; x_\mu (\tau)
\Bigr \}  \nonumber\\
&& \bigl [ X_\mu (\tau), P_\nu (\tau) \bigr ]\; = \;
(1 + \beta)\; \Bigl \{ i \delta_{\mu\nu}\;
(1 - \beta)\; - \; [x_\mu (\tau), \beta]\; p_\nu (\tau)\Bigr \} \nonumber\\
&& \bigl [ P_\mu (\tau), E (\tau) \bigr ] = 0\;\;\;\;\;\qquad \;\;\;\;
\bigl [ E (\tau)  (\tau), E (\tau) \bigr ] = 0 \nonumber\\
&& \bigl [ X_\mu (\tau), E (\tau) \bigr ]\; = \; 2\; (1 + \beta)\;
[ x_\mu (\tau), \beta ]\; e (\tau)\;\; \qquad
\bigl [ P_\mu (\tau), P_\nu (\tau) \bigr ] = 0.
\end{array} \eqno(3.5)
$$
A few comments, at this juncture, are in order now. First,
it is obvious from the above that the limit $\beta \to 0$ produces the
canonical brackets (3.4) for the massless free relativistic particle.
Second, the commutator between the transformed momentum fields
as well as the einbein fields remains the same as their untransformed
canonical form. Third, the spacetime becomes noncommutative very naturally
due to the new symmetry transformations (3.2) (and its infinitesimal 
version(3.3)). This should be contrasted with the NC that
emerges due to the choices of the gauge in the context of the
Dirac bracket formalism (see, e.g., [17,18]). Fourth,
it will be noted that the brackets in (3.5) are still up to the  order 
$\beta^2$. However, for the discussion of the dynamics in the next section, 
we shall be concentrating only on the contributions coming from the
brackets of order $\beta$. Finally, for one of the simplest choices:
$\beta (e^2p^2) = e^2 p^2$, the above brackets (3.5) become
$$
\begin{array}{lcl} 
&& \bigl [ X_\mu (\tau), X_\nu (\tau) \bigr ]\; = \; - 2 i e^2
(1 + e^2 p^2) \Bigl \{ (x_\mu (\tau) p_\nu (\tau)
- x_\nu (\tau) p_\mu (\tau) \Bigr \} 
\equiv - 2 i e^2 M^{*}_{\mu\nu} \nonumber\\
&& \bigl [ X_\mu (\tau), P_\nu (\tau) \bigr ]\; = \;
(1 + e^2 p^2) \Bigl \{ i \delta_{\mu\nu}\;
(1 - e^2 p^2) - 2 i e^2 p_\mu\; p_\nu \Bigr \} \nonumber\\
&& \bigl [ P_\mu (\tau), E (\tau) \bigr ] = 0\;\;\; \;\;\;\qquad\;\;\;
\bigl [ E (\tau), E (\tau) \bigr ] = 0 \nonumber\\
&& \bigl [ X_\mu (\tau), E (\tau) \bigr ]\; = \; 4\;i\;e^3\; (1 + e^2 p^2)\;
p_\mu\; \qquad
\bigl [ P_\mu (\tau), P_\nu (\tau) \bigr ] = 0
\end{array} \eqno(3.6)
$$
where $M_{\mu\nu}^* = (1 + e^2 p^2) M_{\mu\nu}$. This shows that (i) the
NC of the spacetime in the above owes its origin to the rotation in the
off-shell (i.e. $p^2 \neq 0$) reference frame
that is generated by the (off-shell) angular momentum operator
$M^*_{\mu\nu} = (1 + \beta) M_{\mu\nu}$
where $\beta = e^2 p^2$. (ii) The antisymmetric
property of the commutator (i.e. $[X_\mu, X_\nu] = - [X_\nu, X_\mu]$)
is encoded in the antisymmetric property of the angular momentum
operator (i.e. $ M^*_{\mu\nu} = - M^*_{\nu\mu}$). (iii) It is evident
that one of the key requirements [10] of the NC 
(i.e. $\int Tr [X_\mu, X_\nu] = 0$) of spacetime geometry is fulfilled, 
in the above, due to the antisymmetric property of $M^*_{\mu\nu}$
(see, e.g., [10]). (iv) It is interesting to check that the
brackets in (3.5) and (3.6) do satisfy all the possible Jacobi identities
among the phase space variables (see, e.g., [16]).\\

\noindent
{\bf 4 Symplectic structures and dynamics}\\

\noindent
For the sake of simplicity, we shall focus on the motion
of the free massless relativistic particle on a two-dimensional 
Euclidean target (i.e. configuration) space
parametrized by the coordinate variables  $x_1 (\tau)$ and $x_2 (\tau)$. The 
corresponding four-dimensional phase (i.e. cotangent) space 
is parametrized by the four variables 
$x_1 (\tau), x_2 (\tau), p_1 (\tau), p_2 (\tau)$ where $p_\mu (\tau)$'s
(with $\mu = 1, 2$)
are the canonical conjugate momenta corresponding to the coordinate
variables $x_\mu (\tau)$'s  $(\mu = 1, 2)$. The following Hamiltonian
function $H (1)$
$$
\begin{array}{lcl} 
H (1) = {\displaystyle \frac{1}{2}\; e (\tau)\; p^2 (\tau)
\equiv \frac{1}{2}\; e (\tau)\;
\bigl [\; p_1^2 (\tau) + p_2^2 (\tau) \; \bigr ]}
\end{array} \eqno(4.1)
$$
with the canonical commutators $ [x_\mu (\tau), p_\nu (\tau) ] 
= i \delta_{\mu\nu},
[x_\mu (\tau), e (\tau)] = 0, [p_\mu (\tau), p_\nu (\tau)] = 0,
[p_\mu (\tau), e(\tau)] = 0, [x_\mu (\tau), x_\nu (\tau)] = 0, $ 
leads to the following equations of motion
$$
\begin{array}{lcl} 
\dot x_\mu (\tau) = -\; i\; \bigl [\; x_\mu (\tau), H(1) \;\bigr ] 
= e(\tau)\; p_\mu (\tau)
\qquad \dot p_\mu (\tau) =
 -\; i\; \bigl [\; p_\mu (\tau), H(1) \;\bigr ] = 0
\end{array} \eqno(4.2)
$$
which imply the validity of (3.1) as well as the free motion  
(i.e. $\dot p_\mu = 0$) of the massless relativistic particle. Classically,
the above canonical commutators correspond to the canonical Poisson brackets
$\{ x_\mu, x_\nu \}_{(PB)} = 0, \{ p_\mu, p_\nu \}_{(PB)} = 0,
\{x_\mu, p_\nu \}_{(PB)} = \delta_{\mu\nu}$ on the four dimensional
symplectic (i.e. cotangent) manifold with the following contravariant
and covariant symplectic structures
$$
\begin{array}{lcl} 
\Omega^{AB} (1) =
\left ( \begin{array}{cccc}
0 & 0 & 1 & 0\\
0 & 0 & 0 & 1\\
-1 & 0 & 0 & 0\\
0 & -1 & 0 & 0\\
\end{array} \right ) \qquad
\Omega_{AB} (1) =
\left ( \begin{array}{cccc}
0 & 0 & -1 & 0\\
0 & 0 & 0 & -1\\
1 & 0 & 0 & 0\\
0 & 1 & 0 & 0\\
\end{array} \right ) 
\end{array} \eqno(4.3)
$$
where ``1'' in the round brackets after $\Omega^{AB}$ and $\Omega_{AB}$
stands for similar round bracket in
the Hamiltonian function (4.1) and the notation
$z^A = (z^1, z^2, z^3, z^4) \equiv  (x_1, x_2, p_1, p_2)$ has been introduced
in the definition of the matrix form of the symplectic structures as
$$
\begin{array}{lcl} 
\Omega^{AB} = \mbox{Matrix} \bigl (\{ z^A, z^B \}_{(PB)} \bigr )\; \qquad
\Omega^{AB} \Omega_{BC} = \delta^A_C = \Omega_{CM} \Omega^{MA}.
\end{array} \eqno(4.4)
$$
The most general form of the Poisson brackets between any two arbitrary
dynamical variables $F(z)$ and $G(z)$ on the symplectic cotangent manifold
is defined by exploiting the contravariant symplectic structure as
$$
\begin{array}{lcl} 
\{ F(z), G(z) \}_{(PB)} = \Omega^{AB}\; \partial_A F(z) \partial_B G(z)\;
\qquad \partial_A = {\displaystyle \frac{\partial} {\partial z^A}}.
\end{array} \eqno(4.5)
$$
On the other hand, the covariant symplectic structure plays a pivotal role
in the definition of the Legendre transformation which leads to the derivation
of the first-order Lagrangian from a given Hamiltonian. In general, the
symplectic structures can be functions of the phase variables $z^A$. In such
a case, the general form of the Legendre transformation is given
by (see, e.g., [25,26] for details)
$$
\begin{array}{lcl} 
L_f (z, \dot z) = z^A \bar \Omega_{AB} (z) \dot z^B - H (z)
\end{array} \eqno(4.6)
$$
where the general form of the covariant symplectic structure 
$\bar \Omega_{AB} (z)$ is [25,26]
$$
\begin{array}{lcl} 
\bar \Omega_{AB} (z) = {\displaystyle \int}_0^1 d \kappa \; \kappa\;
\Omega_{AB} (\kappa z).
\end{array} \eqno(4.7)
$$
For our present case of the symplectic structures, defined in (4.3)
and satisfying (4.4), the above formula yields $\bar \Omega^{AB} (1)
= \frac{1}{2}\; \Omega_{AB} (1)$. Modulo some total derivatives with respect
to $\tau$, the above equation (4.6) for the Legendre transformation,
produces the following first order Lagrangian
$$
\begin{array}{lcl}
L_f (1) = p_\mu (\tau) \dot x^\mu (\tau) - \frac{1}{2} \; e (\tau) p^2 (\tau)
\end{array} \eqno(4.8)
$$
which is exactly the same as in (2.2)  for the massless relativistic particle.
The equations of motion from the above Lagrangian lead to the derivation of
exactly the same equations of motion as given in (3.1) and the substitution 
$p_\mu = e^{-1} \dot x_\mu$ in (4.8) produces $L_s$ of (2.2).

We shall now exploit the above simple discussion in the context of the
noncommutative commutators given in (3.5) (and its special case given in
(3.6)) for the choice $\beta = e^2 p^2$. First of all, using the infinitesimal
transformations in (3.3), we obtain the transformed version of the 
Hamiltonian function $H (2)$ from its untransformed version $H (1)$, as
$$
\begin{array}{lcl} 
H(1) = \frac{1}{2}\; e (\tau) \; p^2 (\tau)\;\;\; \Rightarrow\;\;
H(2) = \frac{1}{2}\; E (\tau) \; P^2 (\tau).
\end{array} \eqno(4.9)
$$
It is interesting to point out that the {\it form} of our beginning equations
of motion, 
derived from $H(1)$ and expressed in terms of the 
untransformed variables, remains {\it unchanged} up to
the lowest order in $\sim e^2 p^2$ when 
the equations of motion, in terms of the transformed variables,
are derived from
the transformed Hamiltonian $H(2)$. This can be checked by
considering the following
Heisenberg's equations of motion for the transformed variables
$$
\begin{array}{lcl} 
\dot X_\mu (\tau) &=& -\; i\; \bigl [\; X_\mu (\tau), H(2) \;\bigr ] 
\equiv  - \frac{i}{2}\; [X_\mu (\tau), E(\tau)] \; P^2 (\tau) \nonumber\\
&-& \frac{i}{2}\; E(\tau)\;
[X_\mu (\tau), P^2 (\tau)] \qquad \;\;\;
\dot P_\mu (\tau) =
 -\; i\; \bigl [\; P_\mu (\tau), H(2) \;\bigr ] = 0
\end{array} \eqno(4.10)
$$
where we have exploited the commutation relations of (3.6) to prove the
free motion $(\dot P_\mu = 0$) of the particle. Furthermore, with the help of
(3.6), it can be seen that the explicit expressions
for the commutators of (4.10), up to the order $\beta = e^2 p^2$, are
$$
\begin{array}{lcl} 
&&- \frac{i}{2}\;[ X_\mu (\tau), E(\tau)] \;P^2 (\tau)
\approx  2 \;(e p_\mu)\; (e^2 p^2) + {\cal O} (e^4p^4) +... \nonumber\\
&& - \frac{i}{2}\; E (\tau)\; [X_\mu (\tau), P^2 (\tau)]
\approx  E(\tau) P_\mu (\tau) - 2\; (e p_\mu)\; (e^2 p^2) +
{\cal O} (e^4p^4).
\end{array} \eqno(4.11)
$$
This demonstrates that, using the non-trivial commutators of (3.6), we
obtain the equations of motion from the transformed Hamiltonian $H(2)$
$$
\begin{array}{lcl} 
\dot X_\mu (\tau) = E (\tau) \; P_\mu (\tau)\;\; \qquad \dot P_\mu (\tau)  = 0
\end{array} \eqno(4.12)
$$
which dynamically corresponds to the same equations of motion as
$\dot x_\mu  = e p_\mu, \dot p_\mu = 0$. It should be noted that the 
equations of motion remain {\it form invariant} only up to the order 
$\beta = e^2 p^2$ of the transformations in (3.2). They do not retain this
form-invariance even at the next order (i.e. $\beta^2 = e^4 p^4$). To obtain
the analogue of (4.8) for the noncommutative brackets (cf. (3.6)), we have to
exploit the analogues of the definitions in (4.6) and (4.7) 
which heavily depend on the explicit form
of the symplectic structures. To this end in mind, we obtain
the Poisson brackets from (3.6) (valid up to the order $\sim e^2p^2$) as
$$
\begin{array}{lcl} 
&& \bigl \{ X_\mu (\tau), X_\nu (\tau) \bigr \}_{(PB)}\; = \; - 2 \; e^2
(1 + e^2 p^2) \Bigl ( x_\mu \;p_\nu 
- x_\nu \; p_\mu  \Bigr ) \nonumber\\
&& \bigl \{ X_\mu (\tau), P_\nu (\tau) \bigr \}_{(PB)}\; = \;
\delta_{\mu\nu}\;
- 2  e^2 (1 + e^2 p^2)\;p_\mu\; p_\nu\; \quad
\bigl \{ P_\mu (\tau), P_\nu (\tau) \bigr \}_{(PB)} = 0.
\end{array} \eqno(4.13)
$$
In more explicit form, the above brackets yield the following brackets
for the noncommutative four-dimensional cotangent manifold parametrized by 
four phase variables
\footnote{It should be noted that an actual computation of the Poisson 
brackets between spacetime variables $X_\mu$'s is :
$ \{ X_\mu (\tau), X_\nu (\tau)  \}_{(PB)} =  + 2 \; e^2
(1 + e^2 p^2)\; (p_\mu x_\nu - p_\nu x_\mu)$. However, to be consistent with
the corresponding commutator in (3.6), we have exploited the substitution
$(p_\mu x_\nu - p_\nu x_\mu) = - (x_\mu p_\nu - x_\nu p_\mu)$ which,
to be very precise, is valid for the usual definition of a
commutator $[x_\mu p_\nu ] = x_\mu p_\nu
-  p_\nu x_\mu = i \delta_{\mu\nu}$. We have followed here the usual
convention that connects a commutator with the corresponding Poisson bracket 
(i.e. $[ F, G] = i \{ F, G \}_{(PB)}$)
for a couple of dynamical
variables $F$ and $G$.}
$$
\begin{array}{lcl} 
&& \bigl \{ X_1, X_1 \bigr \}_{(PB)} = 0\; \qquad
 \bigl \{ X_2, X_2 \bigr \}_{(PB)} = 0 \nonumber\\
&&\bigl \{ X_1, X_2 \bigr \}_{(PB)} = J_{12} (z)\; \qquad
\bigl \{ X_2, X_1 \bigr \}_{(PB)} = - J_{12} (z) \nonumber\\
&& J_{12} (z) = - 2 \; e^2\; (1\; +\; e^2 \; p^2)\;
(x_1\; p_2 - x_2 \; p_1)
\end{array} \eqno(4.14)
$$
$$
\begin{array}{lcl} 
&& \bigl \{ X_1, P_1 \bigr \}_{(PB)} =  1 - 2 e^2 (1 + e^2 p^2)\; p_1^2
\equiv S_{11} (z)
\nonumber\\
&& \bigl \{ X_2, P_2 \bigr \}_{(PB)} =  1 - 2 e^2 (1 + e^2 p^2)\; p_2^2
\equiv S_{22} (z) \nonumber\\
&& \bigl \{ X_1, P_2 \bigr \}_{(PB)} =  - 2 e^2 (1 + e^2 p^2)\; p_1 p_2
\equiv S_{12} (z) \nonumber\\
&& \bigl \{ X_2, P_1 \bigr \}_{(PB)} =  - 2 e^2 (1 + e^2 p^2)\; p_2 p_1
\equiv S_{12} (z) \nonumber\\
&& \{ P_1, P_1 \}_{(PB)} = 0\; \qquad
 \{ P_2, P_2 \}_{(PB)} = 0 \nonumber\\
&&\{ P_1, P_2 \}_{(PB)} = 0\; \qquad
\{ P_2, P_1 \}_{(PB)} = 0 
\end{array} \eqno(4.15)
$$
where now the symbol $z^A$ stands for: $z^A = (z^1, z^2, z^3, z^4)
= (X_1, X_2, P_1, P_2)$. 
According to our definition in (4.4), we obtain the following contravariant
symplectic structure
$$
\begin{array}{lcl} 
\Omega^{AB}_{(2)} (z) =
\left ( \begin{array}{cccc}
0 & J_{12} (z) & S_{11} (z) & S_{12} (z)\\
-J_{12} (z) & 0 & S_{12} (z) & S_{22} (z)\\
-S_{11} (z) & -S_{12} (z) & 0 & 0\\
-S_{12} (z) & -S_{22} (z) & 0 & 0\\
\end{array} \right ). 
\end{array}\eqno(4.16)
$$
The covariant symplectic structure, corresponding to
the above contravariant symplectic structure and satisfying (4.4), is
$$
\begin{array}{lcl}
\Omega_{AB}^{(2)} (z) = {\displaystyle \frac{1}{S_{12}^2 (z) 
- S_{11} (z) S_{22} (z)}}\;
\left ( \begin{array}{cccc}
0 & 0 & S_{22} (z) & -S_{12} (z)\\
0 & 0 & -S_{12} (z) & S_{11} (z)\\
-S_{22} (z) & S_{12} (z) & 0 & -J_{12} (z)\\
S_{12} (z) & -S_{11} (z) & J_{12} (z) & 0\\
\end{array} \right ). 
\end{array} \eqno(4.17)
$$
Exploiting the definition  (4.7), we
compute the covariant symplectic metric $\bar \Omega^{(2)}_{AB} (z)$ which will
be useful in the context of the Legendre transformations (4.6). It is
elementary to check that $\bar \Omega^{(2)}_{ij} (z) = 0$ for $i, j = 1, 2$.
The next non-trivial component of the covariant symplectic metric is
$$
\begin{array}{lcl} 
\bar \Omega^{(2)}_{13} (z) = {\displaystyle \int}_{0}^{1}\;d \kappa\; \kappa\;
\Bigl [\;
{\displaystyle \frac{S_{22} (\kappa z)}{S_{12}^2 (\kappa z) - S_{11} (\kappa z)
S_{22} (\kappa z)}}\; \Bigr ]
\end{array} \eqno(4.18)
$$
where by explicit computation, it can be seen that, to the order 
$\sim e^2 p^2$, we have
$$
\begin{array}{lcl} 
&&S_{22} (\kappa z) \approx (1 - 2 e^2 p_2^2 \kappa^2)\; \qquad
S_{12}^2 (\kappa z) \approx  4 \; (e^4 p_1^2 p_2^2)\;\kappa^4 \approx 0
\nonumber\\
&&S_{11} (\kappa z) \approx (1 - 2 e^2 p_1^2 \kappa^2)\; \qquad
S_{11} (\kappa z) S_{22} (\kappa z) \approx ( 1 - 2 e^2 p^2 \kappa^2).
\end{array} \eqno(4.19)
$$
Insertions of these values in (4.18) yield
$$
\begin{array}{lcl} 
\bar \Omega^{(2)}_{13} (z) 
= - {\displaystyle \int}_{0}^{1}\;d \kappa\; \kappa\;
\Bigl [\;
{\displaystyle \frac{ (1 - 2 e^2 p_2^2 \kappa^2)}
{ ( 1 - 2 e^2 p^2 \kappa^2)}}\; \Bigr ]
\approx - {\displaystyle \frac{1}{2}} (\; 1 + e^2 p^2\; )
\end{array} \eqno(4.20)
$$
where we have used the following integrals
$$
\begin{array}{lcl} 
&&- {\displaystyle \int}_{0}^{1}\;
\Bigl [\;
{\displaystyle \frac{d \kappa \; \kappa}
{ ( 1 - 2 e^2 p^2 \kappa^2)}}\; \Bigr ] = 
{\displaystyle \frac{1}{4 e^2 p^2}}\; {\displaystyle \int}_{1}^{1 - 2 e^2 p^2}
{\displaystyle \frac{dt}{t} }\;
\equiv  - {\displaystyle \frac{1}{2}}\; (1 + e^2 p^2) 
+ {\cal O} (e^4 p^4) + ... \nonumber\\
&&+ 2 e^2 p_2^2\;{\displaystyle \int}_{0}^{1}\;
\Bigl [\;
{\displaystyle \frac{d \kappa\; \kappa^3}
{ ( 1 - 2 e^2 p^2 \kappa^2)}}\; \Bigr ] = 
- {\displaystyle \frac{p_2^2}{4 e^2 p^2}}\; 
{\displaystyle \int}_{1}^{1 - 2 e^2 p^2}
\Bigl [\; {\displaystyle \frac{dt}{t}} - d t\; \Bigr ]
 \equiv  0 + {\cal O} (e^4 p^4).
\end{array} \eqno(4.21)
$$
Thus, it is now evident that, up to the order $\sim e^2 p^2$, we have
$$
\begin{array}{lcl} 
\bar \Omega^{(2)}_{13} (z) = \bar \Omega^{(2)}_{24} (z) = - \frac{1}{2}\;
 (1 + e^2 p^2) \; \qquad
\bar \Omega^{(2)}_{31} (z) = \bar \Omega^{(2)}_{42} (z) = + \frac{1}{2} \;
(1 + e^2 p^2).
\end{array} \eqno(4.22)
$$
Another interesting component of the covariant symplectic metric is
$$
\begin{array}{lcl} 
\bar \Omega^{(2)}_{14} (z) 
= - {\displaystyle \int}_{0}^{1}\;d \kappa\; \kappa\;
\Bigl [\;
{\displaystyle \frac{S_{12} (\kappa z)}{S_{12}^2 (\kappa z) - S_{11} (\kappa z)
S_{22} (\kappa z)}}\; \Bigr ] \approx 
+ {\displaystyle \int}_{0}^{1}\;d \kappa\; \kappa\;
\Bigl [\;
{\displaystyle \frac{S_{12} (\kappa z)}{ (1 - 2 e^2 p^2 \kappa^2)}}\; 
\Bigr ]
\end{array} \eqno(4.23)
$$
where, up to the 
order $\sim e^2 p^2$, $S_{12} (\kappa z) \approx - 2 e^2 p_1 p_2
\kappa^2$. Ultimately, the above integral, with the substitution
$ 1 - 2 e^2 p^2 \kappa^2 = t$,  reduces to
$$
\begin{array}{lcl} 
\bar \Omega^{(2)}_{14} (z) = - 2 e^2 p_1 p_2\;
{\displaystyle \int}_{0}^{1}\;d \kappa\; 
\Bigl [\;
{\displaystyle \frac{ \kappa^3}{ (1 - 2 e^2 p^2 \kappa^2)}}\; 
\Bigr ] \approx {\displaystyle \frac {p_1 p_2}{2 e^2 p^4}}\;
{\displaystyle \int}_{1}^{1- 2 e^2 p^2}\;
\Bigl [\;
{\displaystyle \frac{ d t}{t}}\; - dt\; \Bigr ]
\end{array} \eqno(4.24)
$$
which yields the value of the integral equal to {\it zero} up to the
order $\sim e^2 p^2$. With this result, it is straightforward to see that
$ \bar \Omega^{(2)}_{14} (z) = \bar \Omega^{(2)}_{41} (z) = 0, \;
\bar \Omega^{(2)}_{23} (z) = \bar \Omega^{(2)}_{32} (z) = 0.$
Only one more useful computation is left over as far as the complete
derivation of the covariant metric $\bar \Omega^{(2)}_{AB} (z)$ is concerned.
This is as follows
$$
\begin{array}{lcl} 
\bar \Omega^{(2)}_{34} (z) = 
- {\displaystyle \int}_{0}^{1}\;d \kappa\; \kappa\;
\Bigl [\;
{\displaystyle \frac{J_{12} (\kappa z)}{S_{12}^2 (\kappa z) - S_{11} (\kappa z)
S_{22} (\kappa z)}}\; \Bigr ] \approx 
+ {\displaystyle \int}_{0}^{1}\;d \kappa\; \kappa\;
\Bigl [\;
{\displaystyle \frac{J_{12} (\kappa z)}{ (1 - 2 e^2 p^2 \kappa^2)}}\; 
\Bigr ]
\end{array} \eqno(4.25)
$$
where, up to the order $\sim e^2 p^2$, we have the following form 
for the $J_{12} (\kappa z)$, namely;
$$
\begin{array}{lcl} 
J_{12} (\kappa z) \approx - 2 e^2 (x_1 p_2 - x_2 p_1)\; \kappa^2
- 2 e^2 (e^2 p^2) (x_1 p_2 - x_2 p_1)\; \kappa^4.
\end{array} \eqno(4.26)
$$
Thus, the integral (4.25) finally looks in the following form
$$
\begin{array}{lcl} 
\bar \Omega^{(2)}_{34} (z) &=& 
- 2 e^2 (x_1 p_2 - x_2 p_1)\;
{\displaystyle \int}_{0}^{1}\;d \kappa\;
\Bigl [\;
{\displaystyle \frac{\kappa^3}{ (1 - 2 e^2 p^2 \kappa^2)}}\; 
\Bigr ] \nonumber\\
&-& 2 e^2 (e^2 p^2)\;(x_1 p_2 - x_2 p_1)\;
{\displaystyle \int}_{0}^{1}\;d \kappa\;
\Bigl [\;
{\displaystyle \frac{\kappa^5}{ (1 - 2 e^2 p^2 \kappa^2)}}\; 
\Bigr ].
\end{array} \eqno(4.27)
$$
From our earlier discussions, it is clear that the first integral will
be zero up to the order $\sim e^2 p^2$. The second integral, with the 
substitution $1 - 2 e^2 p^2 \kappa^2 = t$, becomes
$$
\begin{array}{lcl} 
{\displaystyle \frac{(x_1 p_2 - x_2 p_1)}{2 p^2}}\;
{\displaystyle \int}_{1}^{1- 2 e^2 p^2}\;
\Bigl [\;
{\displaystyle \frac{dt}{t}}\; - 2 dt + t\; dt 
\Bigr ]
\end{array} \eqno(4.28)
$$
which is equal to zero up to the order $\sim e^2 p^2$. This finally implies
that, up to the order $\sim e^2 p^2$, we have:
$\bar \Omega^{(2)}_{34} (z) = \bar \Omega^{(2)}_{43} (z) = 0.$
With the help of the above inputs, the covariant symplectic metric
$\bar \Omega^{(2)}_{AB} (z)$, useful for the Legendre transformation (4.6), 
becomes
$$
\begin{array}{lcl} 
\bar \Omega^{(2)}_{AB} (z) = {\displaystyle \frac{1}{2}}\;
\left ( \begin{array}{cccc}
0 & 0 & - (1+e^2 p^2) & 0\\
0 & 0 & 0 & -(1+e^2 p^2)\\
1 + e^2 p^2 & 0 & 0 & 0\\
0 & 1 + e^2 p^2 & 0 & 0\\
\end{array} \right ). 
\end{array} \eqno(4.29)
$$
Exploiting the above covariant symplectic metric, we obtain the first-order
Lagrangian $L_f (2)$, using equation (4.6), as given below
$$
\begin{array}{lcl} 
L_f (2) &=& \frac{1}{2}\; p_1\; (1 + e^2 p^2)\; \dot x_1
+ \frac{1}{2}\; p_2\; (1 + e^2 p^2)\; \dot x_2
- \frac{1}{2}\; x_1\; (1 + e^2 p^2)\; \dot p_1 \nonumber\\
&-& \frac{1}{2}\; x_2\; (1 + e^2 p^2)\; \dot p_2 - \frac{1}{2}\;
e\; \bigl (p_1^2 + p_2^2 \bigr )
\end{array} \eqno(4.30)
$$
where we have used the fact that, up to the order $\sim e^2 p^2$, the 
Hamiltonian functions $H(1)$ and $H(2)$ are one and the same. Now the stage
is set to recall the dynamical restrictions of section 3 where we have
imposed $\dot e = 0, (\dot x \cdot \ddot x) = 0$ (which is also
equivalent to $\dot e = 0, (p \cdot \dot p) = 0$). Tapping this information,
the above first-order Lagrangian (with
$\dot X_1 = (1 + e^2 p^2)\; \dot x_1, \dot X_2 
= (1 + e^2 p^2)\; \dot x_2, X_\mu = (1 + e^2 p^2) x_\mu 
\equiv  x_\mu (1 + e^2 p^2)$) 
can be recast into
$$
\begin{array}{lcl} 
\tilde L_f (2) = \frac{1}{2}\; p_1 \;\dot X_1
+ \frac{1}{2}\; p_2 \;\dot X_2
- \frac{1}{2}\; X_1 \;\dot p_1
- \frac{1}{2}\; X_2 \; \dot p_2 - \frac{1}{2}\;
e\; \bigl (p_1^2 + p_2^2 \bigr )
\end{array} \eqno(4.31)
$$
which leads to the derivation of the following Euler-Lagrange equations
of motion
$$
\begin{array}{lcl} 
\dot X_\mu (\tau) = e p_\mu \;\;\;\;
\qquad \;\;\;\;\;\dot p_\mu = 0.
\end{array} \eqno(4.32)
$$
The above equations are same as the equations of motion (4.12)
(with a bit changed notations).
Thus, the first-order Lagrangian that emerges from
the metric (4.29), is same as (4.8) (modulo the fact that $x_\mu$'s
are now replaced by $X_\mu$)
and therefore the form of the equations
of motion $ \dot x_\mu = e p_\mu$ and $ \dot p_\mu = 0$ remain unchanged
up to the order $\sim e^2 p^2$ as they can be also written as
$\dot X_\mu = E P_\mu$ and  $\dot P_\mu = 0$. In other words, the dynamics
\footnote{ In particular, it can be checked explicitly
that $H (2) = \frac{1}{2} E P^2 \rightarrow H(1) = \frac{1}{2} e p^2$
under transformations  (3.3) up to the order $\sim e^2 p^2$. Thus, despite the
presence of NC in (3.6), the dynamics remains unchanged.} remains
unchanged up to the order $\sim e^2 p^2$. This feature
is exactly same as our earlier
discussion on the Landau problem where, despite the presence of the
NC, the equations of motion for the
charged particle under the influence of the magnetic field, remain
unchanged [27].\\

\noindent
{\bf 5 Connection with quantum groups}\\

\noindent
First of all, let us recapitulate some of the pertinent points
of our earlier work [22] related to the construction of a consistent dynamics
on a four dimensional
noncommutative cotangent manifold. In this connection, it can
be checked that (i) the ordinary Lorentz 
invariance, and (ii) a particular
(i.e. $pq = 1$) quantum group $GL_{q,p} (2)$
invariance are respected together for any arbitrary
ordering of the indices $\mu, \nu$ (with $ \mu, \nu = 1, 2 $) 
in the following relationship between the phase variables
on the cotangent manifold [22] 
$$
\begin{array}{lcl} 
x_\mu x_\nu = x_\nu x_\mu \;\qquad p_\mu p_\nu = p_\nu p_\mu\; \qquad
x_\mu p_\nu = q\; p_\nu x_\mu.
\end{array} \eqno(5.1)
$$
The phase variables $x_\mu (\mu = 1, 2)$ and 
$p_\mu (\mu = 1, 2)$, in the above,  undergo the following change
under the quantum group $GL_{q,p} (2)$ transformations
$$
\begin{array}{lcl} 
\left ( \begin{array}{c}
x_1\\
p_1\\
\end{array} \right ) \rightarrow
\left ( \begin{array}{c}
X_1\\
P_1\\
\end{array} \right ) &=&
\left ( \begin{array}{cc}
A & B\\
C & D\\
\end{array} \right ) 
\left ( \begin{array}{c}
x_1\\
p_1\\
\end{array} \right ) \nonumber\\
\left ( \begin{array}{c}
x_2\\
p_2\\
\end{array} \right ) \rightarrow
\left ( \begin{array}{c}
X_2\\
P_2\\
\end{array} \right ) &=&
\left ( \begin{array}{cc}
A & B\\
C & D\\
\end{array} \right ) 
\left ( \begin{array}{c}
x_2\\
p_2\\
\end{array} \right ) 
\end{array} \eqno(5.2)
$$
where the elements $A, B, C, D$ of the $2 \times 2$ quantum 
matrix belonging to the quantum group 
$GL_{q,p}(2)$ obey the braiding relations in rows and columns
as 
$$
\begin{array}{lcl} 
&& A B = p B A\; \qquad A C = q C A\; \qquad B D = q D B\; \qquad
B C = (q/p) C B \nonumber\\
&& C D = p D C\; \qquad A D - D A \;=\; (p - q^{-1})\; B C 
= ( q - p^{-1})\; C B.
\end{array} \eqno(5.3)
$$
It will be noted that the noncommutative
algebraic relations (5.1) remain form-invariant
(i.e. $X_\mu X_\nu = X_\nu X_\mu, P_\mu P_\nu = P_\nu P_\mu,
X_\mu P_\nu = q P_\nu X_\mu$)
under (5.2) only for the non-zero complex 
deformation parameters $q, p$
(i.e. $q, p \in {\cal C}/\{0\}$)
satisfying the restriction : $ p q = 1$. In other words, the
quantum group $GL_{q, q^{-1}} (2)$ is responsible for the form-invariance 
of the relations (5.1) on the cotangent manifold. For this group, the
$q$-algebraic relations (5.3) among the elements $A, B, C, D$
reduce to the following simpler form
$$
\begin{array}{lcl} 
&& A B = q^{-1} B A\; \qquad A C = q C A\; \qquad B D = q D B \nonumber\\
&& B C = q^2 C B\; \qquad
C D = q^{-1} D C\; \qquad A D = D A.
\end{array} \eqno(5.4)
$$
At  this stage, two comments are in order. First,
for our present discussion about the dynamics, 
we have chosen the four dimensional
Euclidean noncommutative cotangent manifold 
{\it only} for the sake of simplicity. Our 
present discussions, however,  can 
be generalized to any arbitrary $2N$-dimensional
($N > 2$) cotangent manifold in a 
straightforward manner. Second,
it will be noted that the elements $A, B, C, D$ of
the quantum group $GL_{q, q^{-1}} (2)$ are assumed to {\it commute}
(i.e. $A x_\mu = x_\mu A, p_\mu A = A p_\mu, A X_\mu = X_\mu A$, etc.) with
the phase variables $(x_\mu, p_\mu)$ and $(X_\mu, P_\mu)$ in the proof of
the form-invariance of (5.1).

Now the stage is set for a thorough discussion on the new scale transformations
for the phase variables:
$ x_\mu \to X_\mu = (1 + e^2 p^2) x_\mu, p_\mu \to P_\mu
= (1 - e^2 p^2)  p_\mu$  of equation (3.3) for the choice 
$ \beta = e^2 p^2$ in the framework 
of the quantum groups. To this end in mind, it can be seen that for
$B = 0, C = 0$ in (5.2) (i) we obtain a scale type of transformations
for the phase variables (i.e. $x_\mu \to X_\mu = A  x_\mu, p_\mu \to P_\mu
= D  p_\mu$), (ii) the noncommutative relations of (5.1) remain
form-invariant for $B = C = 0$, and (iii) the algebraic relations (5.4)
reduce to a single relationship $ A D = D A$. In all the above  observations,
we have assumed the {\it commutativity} of the elements $A, D$ with the phase
variables $x_\mu$ and $p_\mu$. This crucial assumption must be {\it maintained}
in our attempt to capture the transformations (3.3) for $\beta = e^2 p^2$
in the framework of the quantum groups. For instance, for the choice
$A = (1 + e^2 p^2)$ and $D = (1 - e^2 p^2)$, it can be seen that
$ A D = D A = 1$ up to the order $\sim e^2 p^2$. Thus, for the above choice
of $A$ and $D$ (with $B = C = 0$), we see that (i) 
the relationship $A D = D A$,
corresponding to the quantum group $GL_{q, q^{-1}} (2)$ is satisfied,
and (ii) the scale transformations $X_\mu = (1 + e^2 p^2)\; x_\mu$
and $P_\mu = (1 - e^2 p^2)\; p_\mu$ of (3.3) (i.e. $\beta = e^2 p^2$)
are captured
in the language of the quantum group transformations. 
Thus, two of the above three basic requirements are readily fulfilled. In 
addition, the condition $AD = DA = 1$ enforces the quantum group 
$GL_{q,q^{-1}} (2)$ to reduce to the quantum group $SL_{q,q^{-1}} (2)$
because the determinant of the matrix becomes {\it one} (i.e. $AD = 1$).
Finally, let us now concentrate 
on the form-invariance (i.e. $X_\mu X_\nu = X_\nu X_\mu, P_\mu P_\nu
= P_\nu P_\mu, X_\mu P_\nu = q P_\nu X_\mu$) of the relationships  in (5.1)
for the choice $ A = (1 + e^2 p^2)$ and $D = (1 - e^2 p^2)$. In this context,
it is pertinent to recall that the following noncommutative $q$-algebraic 
relations [22]
$$
\begin{array}{lcl} 
&& \dot x_\mu x_\nu = x_\nu \dot x_\mu\; \qquad \dot x_\mu \dot x_\nu
= \dot x_\nu \dot x_\mu\; \qquad \dot x_\mu p_\nu = q p_\nu \dot x_\mu\;
\qquad p_\mu p_\nu = p_\nu p_\mu \nonumber\\
&& e \dot x_\mu = q \dot x_\mu e\; \quad e p_\mu = q p_\mu e\; \quad
e x_\mu = q x_\mu e\; \quad x_\mu x_\nu = x_\nu x_\mu\;  \quad
x_\mu p_\nu =  q p_\nu x_\mu
\end{array} \eqno(5.5)
$$
are valid for the massless $q$-deformed free relativistic particle described 
by the Lagrangian $ L = \frac{1}{2} q^{-1} e^{-1} \dot x^2$ and
the Hamiltonian $H = \frac{1}{2} e p^2$ [22]. The above noncommutative 
$q$-algebraic relations 
have been derived from the consistent differential calculi developed
on the $GL_{q, q^{-1}} (2)$ invariant quantum hyperplane. 
It can be checked that the 
form-invariance of (5.1) could be maintained for our choice
$ A = (1 + e^2 p^2)$ and  $D = (1 - e^2 p^2)$ {\it only} for the restriction
$q^2 = 1$. In fact, this latter restriction emerges from the requirement
$X_\mu P_\nu = q P_\nu X_\mu$ when we exploit
the relations (5.5) that lead to
$$
\begin{array}{lcl} 
p_\mu D = (1 - q^{-2} e^2 p^2)\; p_\mu\;\;\; \qquad\;\;
p_\mu A = (1 + q^{-2} e^2 p^2)\; p_\mu.
\end{array} \eqno(5.6)
$$
Basically, in the above, the restriction on the deformation parameter ($q$) 
emerges because of our
demand that the {\it commutativity} requirement of our earlier discussion
(i.e. $ p_\mu D = D p_\mu, p_\mu A = A p_\mu$) should be true
even if $A$ and $D$ are chosen to be explicitly dependent on the phase 
variables. Such a requirement is very much essential because the derivation
of all the $q$-algebraic relations in (5.5), the definition of the
$q$-deformed Poisson bracket (see, eqn. (5.7) below), etc., are based on
the assumption that the elements of the $GL_{q, q^{-1}} (2)$ group commute
with the physical phase variables $x_\mu$ and $p_\mu$.
It should be emphasized that the restriction $q^2 = 1$ has also appeared in
the context of Landau problem [27] and the requirement of the
equivalence between the gauge 
symmetry and reparametrization symmetry for the case of a
 $q$-deformed relativistic (super)particle [19-21]. 
It is interesting to point out 
that the relations $X_\mu X_\nu = X_\nu X_\mu, 
P_\mu P_\nu = P_\nu P_\mu $ remain sacrosanct for the choice
$A = (1 + e^2 p^2), D = (1 - e^2 p^2)$ for any arbitrary value of $q$
if we use the noncommutative algebraic relations (5.5). This happens
primarily due to  the natural commutativity 
$x_\mu A = A x_\mu, x_\mu D = D x_\mu$ of $x_\mu$ with $A$ and $D$
as well as the commutativity
 $(1 - e^2 p^2) (1 - q^{-2} e^2 p^2) = (1 - q^{-2} e^2 p^2) (1 - e^2 p^2)$.
The restriction $q^2 = 1$ also emerges from the equality of the Poisson
brackets (4.13) with the Poisson brackets calculated from the consideration
of the quantum group $GL_{q, q^{-1}} (2)$. The definition of 
a  Poisson bracket between two
dynamical variables $F (x, p)$ and $G (x, p)$ on the $GL_{q, q^{-1}} (2)$
invariant cotangent manifold, is [22]
$$
\begin{array}{lcl} 
\Bigl \{ F, G \Bigr \}^{(q)}_{PB} = {\displaystyle 
\frac{\partial G} {\partial p_\lambda} \; 
\frac{\partial F}{\partial x_\lambda} - q\;
\frac{\partial G} {\partial x_\lambda} \; 
\frac{\partial F}{\partial p_\lambda}}
\end{array} \eqno(5.7)
$$
where 
(i) the repeated index is assumed to be summed over (i.e. $\lambda = 1, 2$), 
and (ii) the key ingredients have been  taken from the differential calculi
developed on the $GL_{q, q^{-1}} (2)$ invariant quantum hyperplane.
It will be noted that, in the above, the derivatives are defined
as the ``left'' derivatives. This amounts to bringing all the 
specific variables to
the left by exploiting the $q$-algebraic relations of (5.5) before the
differentiation w.r.t. that specific variable could be carried out.
Exploiting the above definition, it can be seen that the following
$q$-deformed noncommutative Poisson brackets, analogous to (4.13), 
emerge in the framework of the quantum group $GL_{q, q^{-1}} (2)$ and 
corresponding differential calculi developed on the 
$GL_{q, q^{-1}} (2)$ invariant hyperplane, namely;
$$
\begin{array}{lcl} 
 \bigl \{ X_\mu (\tau), X_\nu (\tau) \bigr \}^{(q)}_{(PB)}\; 
&=& \; + 2 q^2 \; e^2\;
\bigl [\; (1 + q^{-2} e^2 p^2)\; p_\mu \; x_\nu \nonumber\\
&-& q\; (1 + q^{-4} e^2 p^2)\; p_\nu\; x_\mu \; \bigr ] \nonumber\\
\bigl \{ X_\mu (\tau), P_\nu (\tau) \bigr \}^{(q)}_{(PB)}\; &=& \;
[ 1 + (1 - q^2) e^2 p^2 ]\;\delta_{\mu\nu}\;
- 2  q^2 e^2 [ 1 + q^{-4} e^2 p^2 ]\;p_\mu\; p_\nu \nonumber\\
\bigl \{ P_\mu (\tau), P_\nu (\tau) \bigr \}^{(q)}_{(PB)} &=& 0.
\end{array} \eqno(5.8)
$$
A few comments, at this juncture, are in order. First, it can be seen that
in the limit $ q\to 1$, we get back our Poisson brackets (4.13). Second,
the key restriction on the deformation parameter $q$, that emerges due to the
equality between (4.13) and (5.8) is, once again, $q^2 = 1$. This is due to
the fact that an extra $q$-factor that appears in the definition  of the
$q$-Poisson bracket in (5.7) as well as in (5.8) (see the second term on 
the r.h.s. of the $q$-bracket $\{ X_\mu, X_\nu \}^{(q)}_{PB}$) is due to the 
choice of the contravariant symplectic metric (see, e.g., [22])
\footnote{The $q$-deformed Poisson bracket, expressed in (5.7),
 can be defined in a more symmetric fashion:
$\{ F, G \}^{(q)}_{(PB)} = q^{-1/2} (\partial G/\partial p_\lambda)
(\partial F/\partial x_\lambda) - q^{+1/2} (\partial G/\partial x_\lambda)
(\partial F/\partial p_\lambda)$. This expression corresponds to the choice of 
a contravariant symplectic metric that differs from the one chosen in [22] 
by a constant factor $q^{1/2}$. In this form of the definition of the Poisson 
bracket, the restriction $q^2 = 1$ becomes more transparent.}. 
Third, to shed some more light on the left-derivative, it can be seen 
that the explicit form of a specific $q$-Poisson bracket is
$$
\begin{array}{lcl}
\Bigl \{ X_\mu (\tau), X_\nu (\tau) \Bigr \}^{(q)}_{(PB)}\; 
&=& {\displaystyle \frac{\partial} {\partial p_\lambda}}
\bigl [ (1 + e^2 p^2)\;x_\nu \bigr ]\;
 {\displaystyle \frac{\partial} {\partial x_\lambda}}
\bigl [ (1 + e^2 p^2)\;x_\mu \bigr ]\; \nonumber\\
&-&\; q\;
 {\displaystyle \frac{\partial} {\partial x_\lambda}}
\bigl [ (1 + e^2 p^2)\;x_\nu \bigr ]\;
 {\displaystyle \frac{\partial} {\partial p_\lambda}}
\bigl [ (1 + e^2 p^2)\;x_\mu \bigr ].
 \end{array} \eqno(5.9)
$$
The meaning of the ``left-derivative'' in the differentiation
$(\partial/\partial p_\lambda) [e^2 p^2]$ is the trick that, using the
$q$-algebraic relations of equation (5.5), the variable $p^2$ should be 
brought to the left. It can be checked that $e^2 p^2 =
q^4 p^2 e^2$ so that $p^2$ is reordered to the left. Now, we apply the
left derivative on it. This operation yields $ 2 q^4 p_\lambda e^2$. This can
be further rearranged to yield $2 q^2 e^2 p_\lambda$. This differentiation
is carried out by exploiting the differential calculi developed in [22].
All the Poisson brackets for the phase variables
in (5.8) have been computed by exploiting the above trick.\\

\noindent
{\bf 6 Conclusions}\\

\noindent
In our present investigation, we have demonstrated the existence of the
noncommutative spacetime structure in the context of a thorough discussion
on the spacetime symmetry properties of the physical system of a free
massless scalar relativistic particle. The presence of an additional
(i.e. new) scale type of spacetime symmetry transformation for this system
entails upon the spacetime to become noncommutative in nature. This new
scale type of spacetime symmetry is drastically different from the usual
scale type of symmetry that belongs to the usual set of conformal group
of spacetime
symmetry transformations. As a consequence, the usual conformal algebra
gets modified and the NC in the spacetime geometry arises
through the noncommutative algebraic structure [16].

It is worthwhile to
compare and contrast the usual scale spacetime symmetry and the additional
scale spacetime symmetry. The key differences are (i) the usual type of
the scale spacetime symmetry is a {\it global} symmetry (cf. (2.3) and (2.4))
but the new scale spacetime symmetry is a {\it local} one {\it per se}.
(ii) The dependence of the local parameter $\beta$ of the additional new scale
type of spacetime symmetry is very specific (i.e. $\beta (\dot x^2)
= \beta (e^2p^2)$) whereas the parameter $\alpha$ of the usual scale spacetime
transformation is global (i.e. spacetime independent). (iii) It is
the requirement of the free ($ \ddot x_\mu = 0, \dot p_\mu = 0$) motion of the
massless relativistic particle that enforces the choice of the gauges
$\dot e = 0, (\dot x \cdot \ddot x) = 0$ (or equivalently
$ \dot e = 0, (p \cdot \dot p) = 0$) which ultimately turn out to be
responsible for the existence of the new scale type spacetime symmetry
(cf. (3.2) and (3.3)). There is no such type of criterion for the existence
of the usual scale type of spacetime symmetry (cf. (2.3) and (2.4)). (iv) It
is the requirement of the consistency and complementarity between the dynamics
and the spacetime symmetries that are at the heart of the existence of the NC
in spacetime structure in our present investigation. This NC is 
{\it intrinsically} different from the NC that arises in the context of the
massive relativistic particle where the choice of gauges leads to
the NC in the spacetime structure in the Dirac bracket formalism [17,18].

One of the central ingredients in our whole discussion is to focus
on the impact
of the NC on the dynamics of the free massless scalar 
relativistic particle. In particular, the discussions on the
equations of motion for this system have been given utmost priority. As it 
turns out, the NC of the spacetime does not affect the
equations of motion up to the order ($\sim e^2 p^2$).
This feature is exactly same as our earlier
work on the Landau problem [27] where the classical equations of motion
remain unaffected by the presence of the NC in the theory.
In our present paper, we have carried out a detailed and systematic computation
for the derivation of the equations of motion from the Lagrangian and 
Hamiltonian formulations. The key role in all these discussions is played
by the contravariant- and covariant symplectic structures which turn out
to be responsible for (i) the systematic
definition of the Poisson brackets on the
cotangent (momentum phase) space, and (ii) the 
consistent definition of the Legendre
transformations, respectively. In particular, section 4 of our present
paper is devoted to a thorough discussion of dynamics where we 
have taken into account the NC of the Poisson brackets
and corresponding symplectic structures. However, 
despite the presence of the parameters of NC in the covariant symplectic
metric (cf. (4.29)), the dynamics
remains unchanged up to the lowest order in the parameter of the NC.

We have attempted to establish a connection between the NC of spacetime
due to the presence of a new scale type of spacetime symmetry and the NC
of spacetime due to the presence of a quantum group $GL_{q,q^{-1}} (2)$
type symmetry for which a consistent dynamics 
has been developed in [22].
In our present investigation, the $2N$-dimensional differential calculi
and the dynamics of [22]
have been reduced to the differential calculi for
a four dimensional cotangent (momentum phase) space where the ordinary
Lorentz invariance and $GL_{q, q^{-1}} (2)$ 
invariance are respected together for any arbitrary ordering of the Lorentz 
indices. The key point, in the above connection between two types of NC, is
the choice of the elements 
(i.e. $B = C = 0, A = (1 + e^2 p^2), D = (1 - e^2 p^2)$)
of the $2 \times 2$ quantum matrix belonging to $GL_{q, q^{-1}} (2)$
which entails upon (i) the spacetime to become noncommutative in nature
(see, e.g., eqns. (5.8), (5.9)),
(ii) the quantum group transformations to capture the new scale 
transformations (3.3) for $\beta = e^2 p^2$, and
(iii) the quantum group symmetry transformations to correspond
to the $SL_{q,q^{-1}} (2)$ transformations because the determinant
of the above $2 \times 2$ quantum matrix becomes one
(i.e. $AD = DA = 1$ for $ B = C = 0$). It should be noted, however,
that the transformations on the einbein field $e(\tau)$ (cf. (3.3)) are
not captured by the quantum group transformations.
This is why the analogues
of the equations of motion (4.10) and (4.32) have not been discussed in the 
framework of the quantum group $SL_{q, q^{-1}} (2)$. Thus, our consideration
of the connection between the NC due to the new scale spacetime symmetry and
the NC due to the quantum group $SL_{q,q^{-1}} (2)$
symmetry is confined only to the analogy between
the Poisson brackets (4.13) and (5.8). As it turns out,
in the limit $q^2 = 1$, the NC in the spacetime
due to the quantum group $SL_{q,q^{-1}} (2)$ symmetry transformations is
reduced to the NC of the spacetime due to the new type of scale symmetry
transformations for the system of a free massless relativistic particle.

It would be interesting endeavour to extend
our present work to the case of a
massless spinning relativistic (super)particle
where the reparametrization and supersymmetric transformations co-exist. It is
expected that the consideration of this system under the
super quantum group $GL_{\surd q} (1|1)$  [20] might turn out, 
at some stage, to be 
quite handy. Furthermore, the noncommutative realization of the
cohomological operators for the super quantum group $GL_{q,q^{-1}} (1|1)$
[28,29] also might play some crucial roles in this context.
We hope to apply our present
work to the case of a massive relativistic particle where some kinds of gauge 
transformations and reparametrization transformations have been shown to
be connected with one-another under a general scheme [17]. These are some
of the issues that are under investigation and our results would be
reported elsewhere [30].\\

\noindent
{\bf Acknowledgement}\\

\noindent
Fruitful suggestions by the referees are gratefully acknowledged.

\end{document}